\documentclass[11pt]{article}
\pdfoutput=1  
\usepackage{graphicx}
\graphicspath{{./figures/}}
\usepackage{appendix}
\usepackage{latexsym,amsmath,amsfonts,amssymb,booktabs}
\usepackage[font=small]{caption}
\usepackage{slashed,upgreek,amscd,cancel,tensor,color}
\usepackage{adjustbox}
\usepackage{epsfig,latexsym,cite}
\usepackage[pdfencoding=auto]{hyperref}
\usepackage{url}
\numberwithin{equation}{section}

\definecolor{MyBlue}{rgb}{0.15,0.15,0.70}

\hypersetup{
colorlinks=true,
citecolor=MyBlue,
linkcolor=MyBlue,
urlcolor=MyBlue
}

\input epsf
\usepackage{amssymb}
\usepackage{amsmath}
\usepackage{amsfonts}
\usepackage{upgreek}
\usepackage{latexsym}
\usepackage{stfloats}
\usepackage{afterpage}
\usepackage{color}

\begin{document}

\begin{center}
\Large{\textbf{(Non)-neutrality of science and algorithms: \\[0.2cm] Machine Learning between fundamental physics and society\footnote{This paper has been published in Italian on the journal \href{http://www.thelabs.sp.unipi.it}{The Lab's Quarterly}. 
The code to reproduce the plots is available at https://github.com/Mik3M4n}  }} \\[1cm]

\large{Aniello Lampo$^{\rm a}$, Michele Mancarella$^{\rm b}$, Angelo Piga$^{\rm c}$}
\\[0.5cm]

\small{
\textit{$^{\rm a}$  Internet Interdisciplinary Institute (IN3), Universitat Oberta de Catalunya, \\ [0.05cm]
Barcelona, Spain}}
\vspace{.2cm}

\small{
\textit{$^{\rm b}$ D\'epartement de Physique Th\'eorique, Universit\'e de Gen\`eve,\\ [0.05cm]
24 quai Ansermet, CH-1211 Gen\`eve 4, Switzerland }}
\vspace{.2cm}

\small{
\textit{$^{\rm c}$ ICFO - Institut de Ci\'encies Fot\'oniques, The Barcelona Institute of Science and Technology, \\ [0.05cm]
Av. Carl Friedrich Gauss 3, 08860 Barcelona, Spain}}
\vspace{.2cm}

\vspace{0.5cm}
\today

\end{center}

\vspace{2cm}

\begin{abstract}
The impact of Machine Learning (ML) algorithms in the age of big data and platform capitalism has not spared scientific research in academia. In this work, we will analyse the use of ML in fundamental physics and its relationship to other cases that directly affect society. We will deal with different aspects of the issue, from a bibliometric analysis of the publications, to a detailed discussion of the literature, to an overview on the productive and working context inside and outside academia.
The analysis will be conducted on the basis of three key elements: the non-neutrality of science, understood as its intrinsic relationship with history and society; the non-neutrality of the algorithms, in the sense of the presence of elements that depend on the choices of the programmer, which cannot be eliminated whatever the technological progress is; the problematic nature of a paradigm shift in favour of a data-driven science (and  society). The deconstruction of the presumed universality of scientific thought from the inside becomes in this perspective a necessary first step also for any social and political discussion. This is the subject of this work in the case study of ML.

\end{abstract}

\newpage

\tableofcontents

\vspace{.5cm}
\section{Introduction}

``It's time to ask: what can science learn from Google?'' asked provocatively Chris Anderson, editor of Wired, in a famous article in 2008 \cite{Anderson2008}. The question implies looking at the core of the scientific method. Building theoretical models, falsifying them, performing tests, is the scheme that every scientist internalise and uses, as a result of centuries of epistemology. Yet, Anderson suggests that, in the petabyte era, this approach is rather out-dated. In his view, given the amount of data available in the 21st century, models are no longer needed to make predictions, thus he declares ``the end of theory". This would represent a ``Copernican revolution'', a paradigm shift with Machine Learning (ML) - namely ``the ability of a machine to learn without being explicitly programmed''  - as the main technical-theoretical pivot. In 2016 alone, almost 10 years after Anderson's article, Google invested between 20 and 30 billion dollars in ML and Artificial Intelligence\footnote{In the literature, several terms are used to describe the possibility of programming a machine to take decisions independently - i.e. without a predictive model. There are subtle differences among them, yet those are often used as synonyms. Specifically, Machine Learning is a branch of Artificial Intelligence, which includes other fields such as Robotics. For sake of clarity, Section II will give a detailed account of other existing technical and terminological differences. In this article, unless otherwise specified, the term Machine Learning will refer to all these techniques.} (AI) - 90\% of which in research and development. Within the same period, the scientific publication rate in this field experienced a dramatic boost\footnote{Data will be discussed in detail in Section II.}.
Physics is no exception. To explain the reasons behind this phenomenon, science's internal criteria - such as technical efficiency and formal convenience - are not sufficient, even if the interest in advanced data mining techniques is far from being surprising. 

This boom of ML in physics is rather the perfect example of the intertwinement between scientific knowledge and the historical reference period. Yet, the dominant position in academia regarding this interconnection can be summarised in a single word: \textit{neutrality}. This would imply that the scientist, and therefore his/her methods as well as the contents he/she produces, would follow a linear progression, disconnected from the social and political context.

On the contrary, this work is based on the assumption that science is, in all respects, a human activity. Science is exercised in a given historical period, within a precise system of thought and production as well as in scientists' everyday life within the framework of a given working relationship. As a consequence, scientific research as a whole necessarily reflects the features of the surrounding world: in this sense it is \textit{non-neutral}. 

The discussion about the non-neutrality of science was at the centre of the epistemological debate between the late nineteenth and twentieth centuries. The debate polarised between the extreme stances of complete independence of science from society, as argued by the Conventionalists \cite{Poincare1905} and the diametrically opposed epistemological anarchism endorsed by Feyerabend \cite{Feyerabend2010}. Several philosophers endorsed more moderate positions, such as Popper \cite{Popper1962}, Lakatos \cite{Lakatos} and Kuhn \cite{Kuhn1962} among the others. In particular, the latter proposed a historical analysis that showed how science develops according to sudden paradigm changes whose affirmation also depends on social factors. 

Taking the use of ML in pure science as a case study, the main purpose of this work is to contextualise the above-mentioned debate in the present. The essay will attempt to demonstrate that, in the field of physics, it is not possible to motivate the use of such tools without taking into account the historical context of reference, namely what is shaping up as ``the age of data and algorithms".

Thus, it is necessary to assess the extent of this trend in physics research and what are the reasons behind it and at the same time to discuss the existing links with the productive and working environments. To this extent, we will speak of ``non-neutrality of science" - as just clarified above - showing how this phenomenon does not depend on dynamics internal to science itself, but rather on a set of economic, social and political factors.

In particular, we will take into account the academic working environment, heavily impacted by the cuts and reforms imposed after the 2008 crisis, resulting in more precarious job arrangements and a significant reduction in permanent positions. In this situation, researchers can, therefore, be expected to select research programs that enable them to acquire skills that can be spent outside of academia rather than based on scientific/research perspectives. Within this framework, the ML boom in academia should be interpreted in light of the great expansion of big data and AI sectors, representing today an industrial sector capable of attracting high-wage workforce and investments.

Indeed, the relationship of ``pure" science with the social and political dynamics can and must be further scrutinised. For instance, \textit{L'Ape e l'Architetto} \cite{Ciccotti1976} highlights how ``pure science's productions play a super-structural role symbolising a specific form of culture''. We refer specifically to the fact that pure sciences\footnote{From here on, with ``pure science'' we will refer to physics and mathematics unless otherwise specified. } contribute to the creation of a system of language, notions, methods and expectations tied to the possibility of describing the entire world in terms of mathematical models\footnote{By ``models'' we refer to predictive models, as in the scientific methodology. In this regard, Anderson's claim aims at replacing models with raw data. The relevance of this aspect will be further discussed.} - even beyond science itself\footnote{A relevant account for this discussion comes from sociology. For instance, Goldthrope promotes a statistical formulation in terms of population theory \cite{Goldthorpe2016}, while Latour \cite{Latour2010} denounces as ``Sociology has been obsessed with the goal of becoming a quantitative science" considering this goal unattainable, if not useless. For a similar critique inspired by Anderson's hypothesis see also \cite{Boyd2011}.}.

This is a key issue nowadays, as ``data", mathematical models and algorithms are often considered as inherently objective elements carrying an impartial ``truth", whose use justifies political decisions as logical consequences of a natural order. Therefore, it is necessary, for any sociological or political discussion around the use of scientific instruments, to deconstruct the claim of universality of the scientific method from inside pure sciences. Hence, the second aim of this work is precisely to start deconstructing the universality and neutrality of ML and AI. We will refer to the ``non-neutrality of the algorithm", meaning the presence within the algorithm itself of elements that require programmers' discretional choices.

If we take Anderson's hypothesis seriously, this argument acquires an even greater significance. By renouncing the theory, namely the construction of falsifiable or improvable predictive frameworks that guide the observations, we are implicitly assuming that ``correlation is enough". This statement is even more problematic in the use of algorithms to regulate social and political life than in epistemological terms. Elevating correlation to a forecasting tool means, in fact, amplifying, strengthening and perpetuating the status quo, essentially the power relationships within society.

The last aim of the work is therefore to problematise the idea of a paradigm shift in favour of a completely data-driven era. On one hand, the analysis of the literature in physics will allow us to assess from the inside whether this is already the case. On the other hand, we have just seen that an explanation exclusively internal to science would not be satisfactory, as the concept of paradigm itself requires putting in relation science to the historical context. In other words, assuming that Anderson's statement is true, we will shed light on how it acquires a different meaning by looking at ``what we can learn from Google", namely the aura of universality and legitimacy that algorithms (and ML in particular) acquire in the era of big data. We will conclude by saying that this scenario puts scientists in front of the duty to assume and state openly that science and its instruments are human activities, thus questionable and never completely neutral. This discussion is a step in this direction through the lens of physics.
 
The article is structured as follows. The first Section will discuss the theoretical framework of this study, i.e. the non-neutrality of science. Section \ref{section2} will introduce the concept of ML and some other technical and historical elements that motivate the surge in its use throughout the past 10 years. In section \ref{section3}, we will then study the use of ML algorithms in physics. First, we will present an estimate of the phenomenon based on bibliometric analysis, and then some concrete cases will be discussed in detail - i.e. high-energy physics \ref{sec:HEP}, astrophysics \ref{sec:astro} and low-energy physics \ref{sec:quant}. It will emerge that both a historical, bibliometric and literature analysis compel us to consider the role of large platforms and the labour market. This will be addressed in Section \ref{section4} . Finally, in Section \ref{section5}  we will be able to develop a parallel between physics and other fields of study and to comment on the concepts of non-neutrality of science and algorithms as well as the issues surrounding the paradigm shift.


\section{The non-neutrality of science }
\label{section2}
By non-neutrality of science we mean the impossibility of a clear demarcation between, on one hand, science's structure - i.e. its theories and internal organization as well as the so-called scientific method -, and, on the other hand, the historical context - namely, the structure of society and its internal power relations.

Many science philosophers focused on the possibility of isolating and defining norms to guide scientists: this is the so-called problem of ``method". Up until today, the objectivity of science is often upheld, along with its value and its social function, based on the assumption that the scientific method is universal. The current scientific paradigm is grounded on this method, which traces back to Galileo and Bacon, passing through Newton and then Einstein, up to modern measurements that finally completed the standard model of elementary particles.
In the Dialogue, Galileo was already aware of the importance of the presence of a speculative theory to guide observation. In fact, he expressed his admiration for scientists who were able to ``prefer what reason told them over that which sensible experiences plainly showed them to the contrary". The standardisation of this lexicon and approach has arguably established through the work of two scientists, Karl Popper and Thomas Kuhn, although many other philosophers - such as Descartes, Locke, Kant, the nineteenth century's ``positivists", the ``conventionalists" of the early twentieth century up to neo-positivists - certainly served as background. 

Notably, in Popper's ``critical rationalism" \cite{Popper1962} science progresses by ``trial and error" - i.e. advancing hypotheses and theories and testing predictions. Yet, such tests can only lead to falsification rather than verification, as scientists cannot be sure that there are other elements, still unknown, that could invalidate the theory. However, the Popperian falsificationist method has had such a strong impact on the epistemological debate that it still defines science. Also, this concept contributes to elevate science above other disciplines, since those do not have the possibility of correcting inherent errors. Behind this notion today hides the alleged objectivity of science. The latter is considered as the prototype of human rational activity, the safest way towards the truth, if not towards certainty, following linear progress. In Popper's vision, ideas change according to criteria that are exclusively internal to science, whose development is therefore cumulative and unrelated to the history of society: in this sense, we could say neutral. 

Therefore, it is imperative to break the barrier between history and science as well as between society and politics on one side, and the linear development of the scientific thought on the other side. It is above all to Lakatos and Kuhn that we owe a turning point in this direction. Every standard or model dominating the scientific culture in history is the product of a ``view of the world" that goes beyond science alone, as it is dictated by a specific historical era with its beliefs and systems of thought. This is what Kuhn defines as a paradigm \cite{Kuhn1962}. Scientific knowledge fluctuates between long periods of ``normal science", consisting in data accumulation according to criteria internal to science and an existing paradigm, and ``revolutions" that lead to the replacement of one paradigm with another, in which the frontier between the ``outside" and the ``inside" of science falls. Several paradigms are \textit{incommensurable} and it is not possible to discard one according to the prescriptions of falsificationism. 
Moreover, experimental data used to logically verify a theory are inevitably contaminated by the assumptions inherent to a given paradigm. Indeed, data themselves are \textit{theory-laden}. 
Paul Feyerabend goes even beyond \cite{Feyerabend2010}. He states that no scientific theory is ever in agreement with all information available in a given time, and ``scientific rationality" standards do not exist if detached from the historical context. The only solution is then to write ``against the method", which means recognizing that standards are a mere ``guide that is part of a guided activity and that is in turn transformed from it". Thus, the osmosis between science, a given era and the underlying political choices is continuous, absolute and inevitable. It is in these terms should talk about science. 
In conclusion, the main outcome is that science must be considered as a human activity, whose development cannot be detached from the historical path traced by social and political factors. In the following sections, this conception of science will be applied to the recent works of theoretical physicists, specifically concerning the increasing use of ML.

\section{From machine learning to deep neural networks}
\label{section3}
In this section, we will introduce several key concepts and definitions that will be useful to frame the issue and the following discussion. 
As already mentioned in the introduction, the term Machine Learning (ML) refers to ``the ability of a machine to automatically learn by itself from data, without being explicitly programmed to do so"\footnote{The introduction of the term Machine Learning is attributed to Arthur Samuel \cite{Arthur1959}. For a historical introduction see the introduction of \cite{Goodfellow2016}. For a formal and rigorous introduction to the issue, refer to  \cite{Mostafa2012} among the others.}. In the 1950s, ML was first employed to explain neurons' functioning using simple binary output models (active/inactive). One of such models is the Perceptron \cite{Rosenblatt1958}. It was clear from the beginning that it was possible to combine several Perceptrons (also called ``neurons") into one layer to provide an output to multiple classes - the ``output layer". This structure is the simplest example of a Neural Network (NN). In NNs, it is possible to progressively combine layers to achieve higher levels of precision and abstraction (we speak of ``hidden layers" for the innermost layers). A NN is an example of a ML algorithm, yet many ML algorithms are not based on neural networks. From now on with ML we will refer to the latter while with NN to the former. 
But it is only in the 2000s that the possibility of developing multi-layers networks has been fully exploited. The reason is threefold. First, the availability of data becomes exponentially larger and varied\footnote{Let's consider some reference datasets: ``Modified National Institute of Standards and Technology" (MNIST) is the 1990s largest dataset amounting to 60,000 handwritten figures. Today the Street View House Numbers (SVHN) - a collection of images of numbers made available by Google Street View - is an order of magnitude larger. The ImageNet dataset has over 14 million images, containing the most diverse objects. Finally, the dataset released by the ``Workshop of Machine Translation" (WMT) reaches 1 billion data.}. Secondly, thanks to the Graphic Processing Units (GPUs) the computing power increases by several orders of magnitude \cite{Raina2009}. These two factors also lead to a third reason which is the need to improve some known algorithms to make them more performing \cite{Hinton2012}-\cite{Maas2013}. 
Thanks to these three key elements (i.e. the volume of data- computing power -new/improved algorithms), we start talking about Deep Learning (DL), or Deep Neural Networks (DNN), referring precisely to the possibility of building networks with many layers. 
The next section will analyse the presence of these three terms in the scientific literature, commenting in detail on some specific cases.

\section{Machine learning in fundamental physics}
\label{section4}
This section will discuss ML's use in research, assessing the extent of this phenomenon and showing the sharp increase of publications in all fields after 2010. It will then focus on physics - mainly on the most significant branches - while analysing in detail several specific cases. 

\subsection{The dramatic increase of scientific publications: data }
\label{sec:data}
In scientific literature, the ArXiv\footnote{www.arxiv.org} stands as the most widely used open-access online archive for sharing articles and preprints. In 2018, 10,000 new articles were posted on this website every month\footnote{https://arxiv.org/stats/monthly\_submissions}. Given its wide use, looking at ArXiv's database is a useful source to get an estimation of this trend. Hence, we have conducted a research of articles containing in the title or the abstract the terms ``Machine Learning'', Neural Networks'' or ``Deep Learning'' between 2005 and 2018. 
Figure 1 below shows the percentage of articles containing the terms ML (red line), DL (green line) and NN (blue line) in the abstract or title. It is immediate to note a sharp rise as of 2014, preceded by a slight increase in 2010. The number of NN-themed articles increased from 113 in 2008 to 4575 in 2017, representing a total increase of 40 times and a relative increase of just over 30. In 2018 alone the number of articles exceeded 8000. Similar estimates apply to the other two terms - i.e. ML and DL - with a relative increase of about 55 times for ML. The term DL is even absent before 2008, while it counts more than 1500 articles in 2017. As can be seen from the graph, the rise in the absolute number of these articles cannot be explained by the general increase in the number of articles submitted to ArXiv, which only doubled, passing from 4970 articles submitted in January 2008 to 10332 articles in December 2017.
\begin{figure}[h]
\centerline{\includegraphics[width=0.9\textwidth]{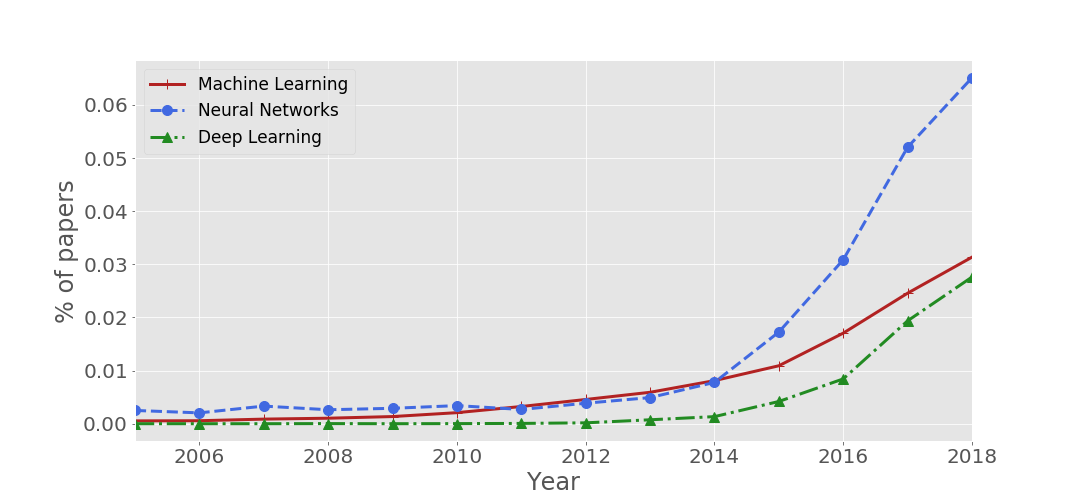}}
\caption{Percentage of articles posted on Arxiv containing ``Machine Learning'', ``Deep Learning'' or ``Neural Networks'' in title or abstract}
\label{fig:fig1}
\end{figure}
It is also interesting to examine which fields of research were most affected by this trend to then focus on physics. ArXiv catalogues articles into 8 categories: Physics, Computer Science, Mathematics, Quantitative Finance, Statistics, Quantitative Biology, Economics, Electrical Engineering and Systems Science.
Figure 2 shows the number of articles on ArXiv containing at least one term among ML, DL and NN in the title or abstract, divided by the above mentioned categories. The data are normalised according to the total number of articles published in the same year within each category and are presented on a logarithmic scale. An exponential growth of research projects with these keywords is therefore clear-cut. As expected, the ``Computer Science" category (red line) dominates the trend in absolute values of published articles, with a significant increase after 2014. ``Statistics" (green line) follows, and right after comes ``Physics" (orange line), whose increase is even more significant (about 10 times between 2008 and 2018). The blue dotted line instead represents the sum of all remaining thematic categories of ArXiv's archive. Even in this case we can observe a significant growth, showing the transversal use of these new techniques in almost all areas of physics.
The period of the surge, particularly in Computer Science, follows a turning point due to the progress in image recognition. In 2012, a DNN won a famous competition, the ImageNet Large Scale Visual Recognition Challenge, by more than 10\% more accuracy \cite{Krizhevsky2012} than the runner-up. This event came to mark the official kick-off of the so-called ``Deep Learning revolution" ultimately convincing major platforms - such as Google, Facebook, Microsoft, etc. - to heavily invest on neural networks' research and development. Our hypothesis, which will be later supported by specific examples, is that physics has been also affected by this surge in investments. 
\begin{figure}[h]
\centerline{\includegraphics[width=0.9\textwidth]{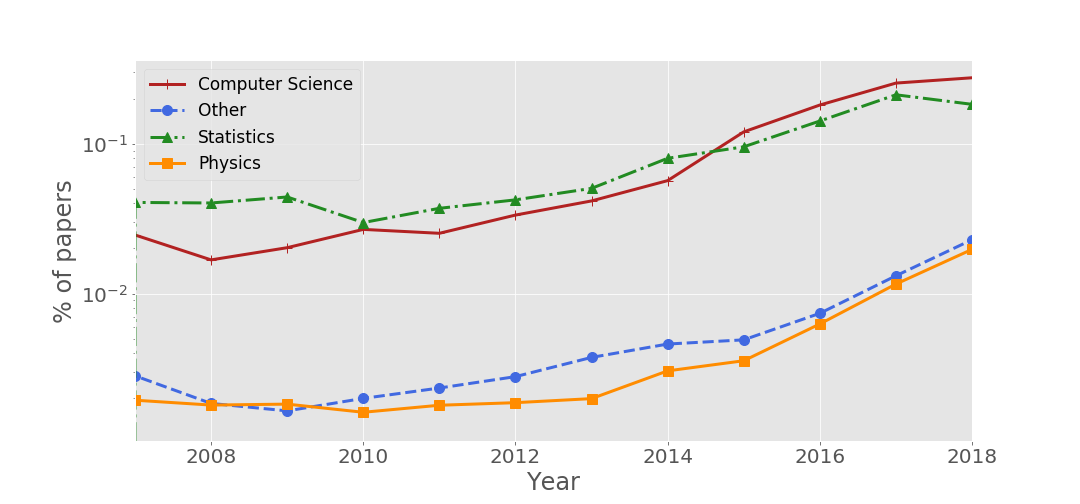}}
\caption{Percentage of articles posted on Arxiv with ``Machine Learning'', ``Deep Learning'' or ``Neural Networks'' in title or abstract divided by the Arxiv section.}
\label{fig:fig2}
\end{figure}
Let's narrow the analysis down to ``Physics''. The latter is divided into 32 different subcategories\footnote{The complete list is available at this link https://arxiv.org/archive/physics }. The graph in Figure 3 shows the percentages of articles in ``Physics'' containing one of the three expressions ML, DL and NN in the title or abstract, divided by subcategories. 
According to the data collected, ``Astrophysics" (astro- ph), ``Condensed Matter Physics" (cond-mat), ``Quantum Physics" (quant-ph) and ``Theoretical High Energy Physics" (hep-th) categories stand out, with a significant post-2014 increase\footnote{To make it easier to read the graph and exclude insignificant information, the graph only displays the subcategories that had an increase of at least a factor 5 between 2008 and 2018, and that had at least 10 articles in 2018.}. Later in the article, each case will be discussed in detail.
It is important to note that this increase involves a wide range of fields, although we will not be able to analyse all cases in detail. The graph summarises all minor fields in a single curve indicated as ``other". This curve seems to indicate, as expected, a growing interest in this topic, specifically where its application is more immediate (such as astrophysics, as we will explain), but also in areas where its application is more pioneering (such as theoretical physics or the diagnosis of tumours in medical physics). In fact, all areas except the first two have almost no publications before 2010. Arxiv's data presented in this section confirm, as we anticipated, the existence of a trend that exceeds significantly the growth in the total number of articles.
\begin{figure}[h]
\centerline{\includegraphics[width=0.9\textwidth]{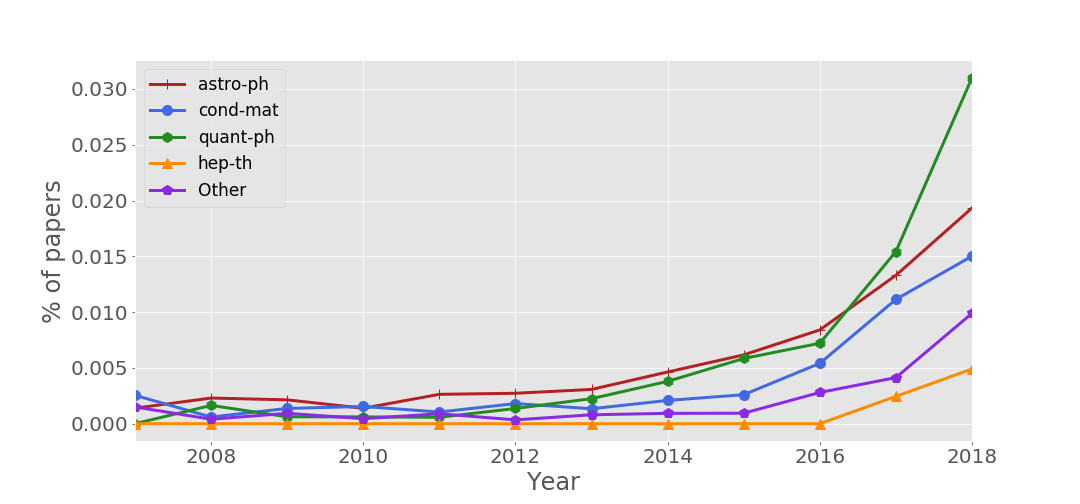}}
\caption{Percentage of Arxiv articles with ``Machine Learning'', ``Deep Learning'' or ``Neural Networks'' in title or abstract in ``physics'', divided by subcategories.}
\label{fig:fig3}
\end{figure}

\subsection{High-energy physics}
\label{sec:HEP}
Statistical techniques and advanced prediction algorithms are fundamental in data analysis. Experimental physics is no exception. In this section, we will discuss one of the most well known collaborations in High-Energy Physics: the Large Hadron Collider (LHC) particle accelerator at CERN in Geneva. 
A review recently published in Nature \cite{Radovic2018} explains how ML has been employed for a long time in LHC; yet, 2014 is marked as a turning point due to the massive investment in the study of DL techniques. The article explicitly mentions Google's advancements in image recognition and the 2012 ImageNet contest among the reasons behind it.  
Still in 2014, the platform Kaggle.com launched a competition, the Higgs Boson Machine Learning Challenge \cite{Adam-Bourdarios2014}. This contest aimed at developing algorithms to reveal the Higgs boson - the famous particle behind the mechanism that gives mass to matter for which the physicist Peter Higgs received the Nobel prize in 2013. This 2014 challenge was won by a combination of 70 neural networks, each with 3 layers of 600 neurons\footnote{https://github.com/melisgl/higgsml/blob/master/doc/model.md}. 
Another key article was published in 2014 officially introducing DL in high-energy physics by comparing it to other ML techniques already in use \cite{Baldi2014}. The authors emphasized the deep conceptual transformation that DL brings, given that ML algorithms work effectively only with data prepared by researchers beforehand based on the knowledge of the physical process involved. This procedure is known as \textit{feature engineering}. For instance, the momentum of a particle revealed at LHC allows reconstructing the ``invariant mass" of the process, a physical quantity with high discriminating power (high-level feature), which is used as input for ML. This step can be extremely complicated or computationally expensive. With a deep neural network, instead, one can directly use the ``raw" data of the detector (low-level features), reaching an equivalent degree of accuracy: the network itself independently retrieves the information that was previously obtained through the knowledge of the physical processes involved.
However, the reverse engineering of a NN, namely the reconstruction of the decisional process that led to the output, is currently impossible. Hence, we give up the knowledge of the physical process in favour of predictive power, as we cannot extract the equivalent of the high-level feature. The above mentioned transition towards a ``blind" analysis is at the core of DL's revolution. This is also the reason why the concept of ``paradigm shift" is so often evoked even in physics recently. In reality, the problem is far more complex as a falsifiable model is still at stake. 
A neural network, in fact, requires the so-called training phase, in which the algorithm learns from a portion of data - the so-called \textit{training set}. In this phase, it is possible to know whether a signal is present or not to then be able to make predictions. The training requires hundreds of thousands of data to be effective. To train the network, the signal sought is then calculated based on the theoretical model that needs to be tested while training data are generated through numerical simulations. In the case of LHC, for instance, the theoretical model to be tested could be the Standard Model - at present the most important and precise model that predicts and catalogues the fundamental particles of nature - or its extensions. At the origin of this process, therefore, remains a predictive model and its ability to provide falsifiable predictions. 
Another pertinent example is the NOvA experiment for neutrinos' detection \cite{Ayres:2007tu}. Specifically, the authors explicitly state that the type of network used is based on the GoogLeNet model \cite{szegedy2014going}, winner of the 2015 ImageNet competition. 
In conclusion, we will mention the study of the quark beauty at LHCb. This is an elementary particle for which models based on the algorithm used by Google translator were recently employed \cite{CERN2017}.

\subsection{Astrophysics}
\label{sec:astro}
Figure 3 shows that astro-ph is one of the subcategories with the highest increase in the number of publications.
This is not surprising given that telescopes provide optical images and image recognition is crucial to effectively extract information. We can look at the detection of strong gravitational lenses\footnote{Strong gravitational lenses are a phenomenon predicted by the theory of General Relativity, which consists in the formation in a telescope of several images originated by the same light source. This is due to the deflection of light originated from gravitational structures along the path it takes to reach the observer's eye.} as an explicit case. At present, the accurate analysis of a single gravitational lens can take even weeks of work and a very thorough knowledge of the physical processes intervening in its observation. A recent article published on Nature \cite{Hezaveh:2017sht} shows that, by using NN\footnote{Once again, one of these networks is AlexNet, winner of the 2012 ImageNet competition.} instead of standard techniques, the detection process is about 10 million times faster and has the same accuracy. Similar to the example of LHC, this approach does not require any specific knowledge of the physical processes involved in the measurement: the data of the telescope's image pass directly to the algorithm. In the past, the measurement required the inclusion of hundreds of thousands of additional parameters to model the signal coupled with complex statistical techniques to extract the final result.
Another example is gravitational-wave detection \cite{Abbott:2016blz}. Recent studies \cite{Gabbard:2017lja} show how it is possible to detect a signal from raw data, achieving the same performance - or even better - compared to the long and complex standard data analysis technique. 
There are also other applications at the frontier of modern astrophysics, such as the study of exoplanets (Lam, 2018; Shallue, 2018) and the next generation cosmological experiments \cite{Gillet:2018fgb}.
The examples discussed - as in the case of high-energies - go in the direction of increasingly data-driven science, compared to a previous era in which the modelling of each component of the experiment was crucial. 
However, even in these cases the training of the network requires training datasets originated from numerical simulations. These, in turn, are issued from complex codes that provide predictions based on the models that are to be tested in the experiment\footnote{In the case of gravitational waves, for example, it is necessary to solve the equations of general relativity for a binary system of compact objects such as black holes or neutron stars, which is the result of numerical techniques combined with accurate analytical approximations. In the case of gravitational lensing, there are galaxy image simulation codes that combine models with existing images.}. In any case, this preliminary process is crucial for using NNs, which are thus reduced to effective tools for data analysis.

\subsection{Quantum mechanics and the physics of condensed matter}
\label{sec:quant}

The last two fields of research on which we will focus on are physics of condensed matter (cond-mat) and quantum mechanics (quant-ph). 
Unlike particle physics and astrophysics, where the examples we have analysed mainly refer to the use of DL techniques for data analysis, quantum mechanics and condensed matter are fields of experimentation for new frontiers. We will list the most important ones\footnote{For a technical review consult \cite{Dunjko2018}.}. We should first mention the so-called quantum machine learning (QML), namely the use of algorithms exploiting the enormous potential of quantum computing to reduce the time and costs to solve ordinary ML problems. QML was born in 2013 through the collaboration between MIT and Google\footnote{See \cite{Lloyd2013}. The authors are Seth Lloyd - global quantum information and quantum computer guru - and the young contributor Patrick Rebentrost, who are both affiliated at MIT, and Masoud Mohseni - senior researcher on the Google AI team. }. In 2013 Google - in collaboration with NASA - purchases a D-Wave quantum computer, justifying the 10 million dollars investment with its potential applications in ML\footnote{Google and NASA Launch Quantum Computing AI Lab: https://www.technology- review.com/s/514846/google-and-nasa-launch-quantum-computing-ai-lab/}. It should be noted that the quantum computer has not been realised yet and QML remains a purely theoretical domain. Though, ML techniques are currently employed to improve experimental processes needed for constructing quantum computers. In this respect, ML is a useful tool rather than a paradigm shift \cite{Biamonte2017}. 
We will conclude this section with a glance at the important applications to the study of quantum phase transitions. Among the traditional problems in this domain there is the study of the properties of atoms - or gas of atoms - interacting at a room temperature often close to zero. The equations needed to reach a complete knowledge of each phase and its dynamic evolution have enormous computational costs and require numerical methods or approximations. Therefore, ML has been used to classify the phases of the system. This approach is very recent as it only appeared in the past 3 years. In \textit{supervised learning} algorithms \cite{Carrasquilla2017} - namely, with an available training set with know phases - researchers already know the system's phases and attempts to identify the exact point of phase's transition - for instance, a given temperature.  
To assess whether we are in front of a paradigm shift, the so-called \textit{unsupervised learning} methods \cite{van_Nieuwenburg_2017} are a far more interesting case, as they do not require prior knowledge of phases. In this case, the algorithm automatically clusters data into affinity groups. However, the feature engineering needed is generally so sophisticated that resorting to ML \cite{Huembeli2018} may no longer be necessary. Another issue is that is often necessary to know in advance how many phases there are in the system. This prevents finding new phases and it requires a qualitative and profound knowledge of the system itself.

\section{Platforms and academia}
\label{section5}
In the previous sections, the link between research and development by large platforms - such as Google - and the subsequent use in physics emerged several times. This suggests the need to investigate beyond physics to reach a sound understanding of the phenomenon. Hence, this section will provide a quantitative measure of the trend of private investments in big data.  An outburst in such investments that is very attractive in the labour market will emerge. In parallel, we will look at the steady deterioration of funds in academia and the shrinking of permanent positions for researchers. We will argue that this scenario pushes young researchers to move towards fields of study that allow them to acquire skills that are suitable for the private sector or basic research. 
Big platforms, the key actors of global capitalism nowadays, base their profit almost exclusively on data, algorithms and computing resources. Almost 95\% of Facebook's profit comes from the sale of advertising space targeted to users' taste\footnote{https://www.statista.com/statistics/267031/facebooks-annual-revenue-by-segment/}, while 80\% of Netflix's streaming comes from recommendations rather than direct research \cite{Gomez-Uribe2016}. Yet, this trend extends to a wide range of sectors that go beyond major platforms. According to a J.P. Morgan's report\footnote{https://flamingo.ai/wp-content/uploads/2017/11/JPMorganAnInvestorsGuideToArtificialIntelligencev2.pdf} the most relevant sectors in which ML is essential are: banks, digital security, finance\footnote{In May 2017, J.P. Morgan published the most extended report ever on Big Data and ML in finance}, governmental use, communication and services, insurances, transports, health and education. 
J.P. Morgan also provides a growth projection of the size of this market, namely 58 billion dollars by 2021 compared to only 12 billion in 2017. This represents one of the fastest-growing technological sectors, with a Compound Annual Growth Rate (CAGR) of almost 50\%. Among the companies surveyed 44\% indicate the ML/AI sector as the technology that will have the greatest impact on the company in the next decade. Based on a different sample O'Reilly Media\footnote{2018 Outlook: Machine Learning and Artificial Intelligence, A Survey of 1,600+ Data Professionals (14 pp., PDF, no opt-in).} estimated a 61\% growth. 
This scenario gives us a picture of the scope of ML's study and applications outside of academia in the past 5-10 years. Likewise, it is possible to find evidence of this trend even in data concerning the labour market. 
According to LinkedIn's 2017 US Emerging Jobs Report published on LinkedIn's website in December 2017\footnote{https://economicgraph.linkedin.com/research/LinkedIns-2017-US-Emerging-Jobs- Report}, the two professions that have seen the highest increase since 2012 are Machine Learning Engineer and Data Scientist, with increases of 9.8 and 6.5 times respectively. Among the first three positions held, before taking on a Machine Learning Engineer or Data Scientist job, we find Research Assistant and Teaching Assistant, giving evidence of the flow of professionals coming from academia. 
Similarly, IBM's 2018 report \textit{The quant crunch: how the demand for data science skills is disrupting the job market}\footnote{https://public.dhe.ibm.com/common/ssi/ecm/im/en/iml14576usen/analytics-analytics- platform-im-analyst-paper-or-report-iml14576usen-20171229.pdf } highlights a 40\% increase in Data Scientist posts in 2016 alone while giving an equal projection of growth by 2020. Among these vacancies, 40\% require a master's degree upwards, while 78\% require at least three years of experience. These data suggest a strong need for high-skilled professionals in scientific disciplines that seem to come mainly from academia. We should now turn our attention to the working conditions in academia. 
Our hypothesis is that much of the interest of young researchers towards ML lies in the possibility of acquiring technical skills suitable for a labour sector that is expanding enormously unlike university where there are little prospects for a permanent appointment, high competition, precariousness and unsustainable stress faced with low wages. 
In the US and Europe, at least until 2008, universities were unable to absorb PhD students and this trend is steadily worsening in all advanced countries\footnote{This does not apply to China, India, Poland and Singapore where public investment in research boomed, for many different reasons, only recently.} \cite{Cyranoski2011}. K. Powell points out that in the US the number of postdocs in science increased by 150\% between 2000 and 2012, without any parallel increase in permanent posts and tenure-track. As a result, either the private sector or the governmental sector - public, but not academic - will absorb PhD graduates. 
Still in the USA, the latest National Science Foundation report, the Science and Engineering indicator 2018\footnote{https://www.nsf.gov/statistics/2018/nsb20181/report}, shows that the percentage of PhDs with chances of getting tenure or tenure-track position is only about 20\%. Since 2000, there has been a sharp decline in full-time faculty posts at the advantage of fixed-term appointments. Narrowing the focus to physical sciences that concern us the most, we observe that, if in 1973 the 21\% of academic positions were filled by physicists, today the percentage has dropped to 14\%. 
If the scarcity of stable positions is likely to force many researchers to leave academia against their vocation, the private sector's high-salaries certainly have a strong attraction. On average an American postdoctoral researcher earns 43,000 dollars at University versus 73,000 dollars as a starting salary in the private sector \cite{Powell2015}. 
Several authors are calling for a drastic reduction of the numbers of PhD combined with a strategic reconfiguration of the structure of research groups and an increase in fixed positions for researchers at the advantage of the productivity ``per capita"  \cite{Gould2015}-\cite{Powell2015}-\cite{Cyranoski2011}. Even if this could appear as a pragmatic argument, given the scarcity of funds, it lacks taking into account the productive context in which university takes place. 
To our understanding, the systemic reason for the high number of postdocs and doctors forced to leave academia is, in fact, the need to train, at the State's expense, an army of highly specialized technicians at the disposal of large and small companies, in other words resources that would be mainly transferred to the private sector. These data should be interpreted in the context of the current labour and productive markets previously outlined. Namely, given the enormous expansion of the AI sector, there is an extremely high demand for Data Science professionals that are filled thanks to the systemic inability of academia to absorb researchers. 
Further confirmation of this trend can be seen the IBM's report mentioned above, which estimates that by 2020 the demand for data technicians will increase by 28\% in the private sector alone, with a peak of 39\% in highly qualified sectors that require a PhD title - where salaries exceed \$100,000. In this regard, IBM denounces a lack of specific profiles and explicitly calls for ``the educational system" to take over their training\footnote{The conclusions of this section are based on data coming from the labour market combined with the analysis of the scientific publications illustrated in the previous sections. It would be useful and necessary to further test these conclusions through an investigation in the academic community, and in particular among young researchers.}.

\section{Machine learning: from physics to society}
\label{section6}
In this section, we will comment on some critical aspects related to ML's use in the social sphere while comparing it to the case of physics. We will highlight that many of these elements are, for intrinsic reasons, part of the process already in pure sciences. 

\subsection{Accuracy, false positives and fairness}
When using mathematical models and tools, a key concept is their ability to provide ``precise" and quantifiable predictions. 
Even without referring to ML, within science itself, the concept as well as the accepted degree of ``accuracy'' varies considerably across different scientific communities. A paradigmatic example dates back to 2015 when the Italian Physical Society (SIF) did not sign the Climate Change Declaration for the Paris Climate Conference COP 21, arguing that the climate studies' accuracy was not sufficient to ``incontrovertibly" affirm the anthropic footprint on climate. Yet, the definition of ``incontrovertibly'' is set by the community together with the threshold beyond which a measure is accepted as evidence, as both are not ``contained'' neither in data itself nor in measurement. These are instead the result of statistical techniques that can vary significantly depending on the field of studies. 
The very definition of ``accuracy" of an algorithm can also vary substantially. Let's investigate the case of ML through an example from physics: a particle detector. The algorithm that regulates the experiment's trigger can be programmed either to minimise the loss of ``authentic signals'' at the expense of having more ``false alarms'' - i.e. events wrongly classified as signal -, or to minimise the ``false alarms'' rate, at the risk of missing some real events. It is important to stress that the \textit{mathematical} definition of these two events (``False positive" and ``False negative") already requires a compromise between the two\footnote{For instance, one can introduce the f-score a weighted average of the two.}. This kind of choices are essential in science, and in the use of algorithms in general: they are part of its nature and cannot be overcome by any technological advancement. If in a given particle physics experiment such decisions could have negligible consequences - e.g. it could take longer to reach a scientific discovery -, we cannot affirm the same concerning a software that has a decisional or advisory role in public life.  This leads us to reformulate these concepts in terms of \textit{fairness} of the algorithm. 
A famous example is the COMPAS\footnote{COMPAS (Correctional Offender Management Profiling for Alternative Sanctions), produced by the Northpointe company, now Equivent (http://www.equivant.com/) } software, used in the United States to predict the risk in releasing a prisoner on bail before trial \cite{Courtland2018}. A team of journalists showed that the software is not ``fair" - contrary to the company's claim - because it predicts more false positives among the African American community. The different definitions of fairness\footnote{In this specific case the predictive parity - i.e. the fact that the probability of being arrested again between white people and black people does not depend on the group to which they belong - and the false positive rate - i.e. the misidentification as a subject at risk. In turn, a minimum false positive rate is irreconcilable with a minimum false negative rate - in this case, misidentification as a subject not at risk.} adopted by the two teams respectively are at the heart of the controversy. This difference is primarily on a mathematical level as different definitions of fairness are statistically incompatible, and the choice of the definition to be used to optimize the algorithm is ultimately in programmers' hands. Professor A. Narayanan, a computer scientist at Princeton, recently held a seminar entitled ``21 fairness definitions and their politic''\footnote{The video-conference is available at \href{https://www.youtube.com/watch?v=jIXIuYdnyyk}{this link}. For the report, see \href{https://docs.google.com/document/d/1bnQKzFAzCTcBcNvW5tsPuSDje8WWWY- SSF4wQm6TLvQ/edit}{here}} showing the many different definitions of this concept that can be applied to an algorithm.
It could be argued that many classifiers have minimal differences - a few percentage points or less - between different types of errors. Yet, when human lives are at stake every detail becomes crucial. This is the case of SKYNET, an algorithm employed by NSA to monitor the Pakistani population by assessing the risk of collusion with terrorism. An information leak\footnote{SKYNET: Applying Advanced Cloud-based Behaviour Analytics. The Intercept, 8 May 2005.} revealed that the level of false positives could range between 0.008\% and 0.18\%. Out of a population of 55 million, this means that up to roughly 100,000 Pakistanis could be erroneously placed under surveillance as suspected terrorists \cite{Hosni2017}. Beyond numbers, this example shows the need to problematize the use of algorithms: whether the error percentage is small or null, is it acceptable to apply a similar tool to monitor an ethnic group?

\subsection{Bias and self-fulfilling hypotheses}
In the first section, we mentioned the concept of ``theory-laden data" introduced by Kuhn. This, in turn, is related to the notion of bias. There are many notorious examples concerning the application of ML algorithms to the social context. Let's turn our attention to one of the most striking ones, namely a study conducted at Stanford University that used a DNN to distinguish heterosexuals and homosexuals through facial recognition \cite{Wang2017}. According to the authors, the accuracy ranged between 80\% and 70\% - far better than human results. Many newspapers have also given their pages over this study, generating a stir as it justified potential neo-Lombrosian hypotheses and genetic causes of homosexuality. A few months later, Google's researchers showed\footnote{The results are available at \href{https://medium.com/@blaisea/do-algorithms-reveal-sexual-orientation-or-just-expose- our-stereotypes-d998fafdf477}{this link}}, with a more sophisticated analysis, that the algorithm was affected by gross gender stereotypes since the training phase. Moreover, they pointed out that, rather distinguishing people based on physiognomy, the algorithm used makeup, glasses and other characteristics regarding people's look and social context rather than genetics. Also, another big issue was at stake, namely that the binary division between heterosexuals and homosexuals is not only controversial but also resulting from stereotypes. This example leaves no space for doubts concerning the non-neutrality of science and algorithms. 
Here, too, we can find analogies with physics. In 1572, Tycho Brahe observed a new very bright celestial body, which he interpreted as the formation of a new star. We currently know that the event was a Supernova Ia, an astronomical event that corresponds to the end of the life cycle of a star, yet unknown at the time of Tycho Brahe. The same classification error could still occur with a modern ML algorithm trained, for instance, to classify the image of a celestial body as ``star" or ``galaxy". Again, the problem at stake is that the number of classes is pre-determined from the outside and guided by theory. 
A further point of reflection comes from the COMPAS software. The fact that the prediction of risk is higher among African Americans and that this is due to a previous bias triggers a process in which such prejudices are perpetuated if not emphasized. 
Another example are the so-called predictive policing software, such as PredPol\footnote{PredPol, Predict Prevent Crime. Predictive Policing Software: www.predpol.com/}, used in the United States to predict the areas where crimes are most likely to occur - without reference to the reason, we stress! The indiscriminate use of available data could create distorted feedback for which, if agents are already in the locations where more crimes have previously occurred, as data suggest, the rate of police interventions in these areas is likely to be higher - regardless the existence of real danger \cite{Ensign2017}. This would turn into a paradoxical vicious circle in which the alleged ``dangerousness" would be confirmed! The fallacy is twofold. First, the assumption that data can reveal valid trends for the future in the absence of an explanation - what is called ``model'' in physics -, and, secondly, the assumption that more police interventions are inevitably due to greater insecurity - a classic case of spurious correlation. But this is nothing new.
In economics this controversy has been running for years: neoclassical economics not only fails its own predictions \cite{SylosLabini2016} but also its tools - obtained within a theoretical paradigm without empirical evidence - lead stakeholders to behave in such a way as to trigger crises \cite{Bouchaud2008}. Once again, it is crucial to emphasise how the presence of a predictive model is understated in the discussion on the role of algorithms. 

\subsection{Data ownership}
In physics, the replicability of experiments and access to data are considered the two inviolable preconditions for drawing any conclusion. 
Moreover, as we have seen, the choice of the type of data to be collected - in other words, the experiment design - is guided by theory and part of the research process. The same cannot be said about modern data warehouses and the use of data-driven models outside academia. Particularly serious is the question of data ownership. If many tools for the extraction of value through predictions are now public\footnote{As it is the case for the python TensorFlow and PyTorch libraries, developed by Google and Facebook respectively to implement deep neural networks.}, the data used to train algorithms represent the real resource for accumulation and these are often private. The joke ``data is the new oil" is very exemplifying in this sense. This phenomenon has an impact on professionals' outflows from academia. Often researchers in public institutions have problems accessing data which is not the case for the big private platforms' employee\footnote{The most striking example is the case of Facebook, which does not allow access to its content via API - not even in a limited form, as in the case of Twitter. For a discussion on some specific cases related to Google and Facebook, see \cite{Cozzo2018}}. The privatisation of data thus generates strict and violent hierarchies, aggravated by the concentration of data ownership in the hands of a few large platforms. 

\subsection{The role of scientists in society}
We have stressed several times how ML algorithms involve in their use a series of steps in which the user is required to make decisions that will affect the output, such as the selection of features or the choice of a probability threshold\footnote{This work focuses on ML, yet this reasoning applies to all kind of algorithms. } and how this arbitrariness is present at various levels. Modern DL algorithms, instead, work as hardly decodable black boxes, in which the process that led to the final decision cannot always be retraced while there are still choices involved, such as setting the criterion of accuracy. In both cases, the possibility of applying a predictive tool that was previously trained on another sample to unknown cases is assumed, completely renouncing to a predictive model. Adopting this ``inductivist" way of proceeding already implies a profound epistemological stance, often ignored outside physics. 
The use of modelling and mathematics has a double implication. On one hand, it turns a social-driven process (or, paraphrasing, ``data-driven") into a falsely objective one (or, one might say, natural, neutral), hence to some extent normalising it. On the other hand, it puts in scientists' hands, through their research, the political decision, raising their role to technician-vehicle of the hegemonic cultural and economic system. If in relation with the first implication, we already warned on the risks of naive applications of mathematical techniques in social sectors by stretching the hypotheses in which they were initially developed (what the domain of applicability is, how the system retroacts, etc.), much more subtle, but equally important and risky, is the second implication. The scientist, in a context of a technical division of labour, is often convinced or pushed to believe, to have a merely technical role and not a political one. Even worst, scientists are sometimes persuaded to have an even higher role as they unveil the truth that is above and beyond politics. A mix of science fetishism, alienation and job insecurity unconsciously strengthens the socio-economic and ideological system, particularly considering that not even scientists themselves are fully aware of algorithms do, given their opacity and bias.

\section{Summary and conclusions}
\label{sec_last}
This work focused on the study of the use of ML in contemporary physics. This choice comes from the awareness of both the inherent historical and political bias of science and its ties with the materiality of its age. Contemporary capitalism is based on the extraction of value from extensive data using advanced ML algorithms and great computing power. These three elements (big data - advanced algorithms - computing power) have marked a turning point in the role of large platforms, thus transforming production and the labour market. This raises promptly the question over the impact of these transformations on science's development. Also, one should investigate what role science has in feeding and legitimizing a certain paradigm in its political and social aspects, and specifically in our case study the key role that algorithms play today. 
The analysis has developed around three elements: 
\begin{itemize}
\item Science is not neutral: scientific research, even in ``pure" sciences, evolves in a non-linear way, tightly following the transformations of society's hegemonic production model. 
\item Algorithms are not neutral, as they are developed in a scientific field that is not neutral and they contain elements that are subject to external decisions and that cannot be eliminated by any technological progress. 
\item In a ``data-driven" paradigm perspective, it becomes all the more important to affirm the impossibility of having objective and neutral instruments even within pure science.
\end{itemize}
We have approached the problem along complementary lines.
Through an historical and bibliometric analysis of the use of ML in physics (section \ref{sec:data}), we showed that there is correlation with technological advances that took place outside academia for reasons related to market needs (extracting value from large bulks of data) that, in turn, determined a profound transformation of production and the labour market. We have also highlighted that the ML boom in physics is also related to scientists' material needs. Subject to precarious working conditions and few career perspectives in academia, researchers are likely to direct their studies and research toward a field of study that is spendable in the labour market (section \ref{section5}).
Secondly, we reviewed the use of ML in the fields of fundamental physics where it is most relevant, namely particle physics, astrophysics, quantum mechanics and condensed matter physics (sections \ref{sec:HEP}-\ref{sec:quant}). It emerged clearly that the above mentioned external factors have a significant impact on academia that can be directly found in the scientific literature. At the same time, we have shown how hypothesis informed by Anderson's idea of ``the end of theory'' - i.e. a science without models - do not hold when confronted with reality, besides any techno-futuristic enthusiasm\footnote{For other critical stances within the scientific community consult \cite{Zdeborova2017}.}.
Finally, we have stressed how problems related to the interpretation of results emerge even in a highly controllable theoretical and experimental context such as that of fundamental physics where variables are presumably well defined. This allows us to extend our criticism to the use of algorithms in the social sphere where bias, prejudice, uncontrollability of the system and feedback assume a direct role and in which people instead of numbers are at stake (section \ref{section6}).
However, looking at the huge impact algorithms have today, their wide use and the increasingly widespread rhetoric based on alleged ``scientific truths", and particularly on data, Anderson's prediction - even if based on false assumptions - could come true.  Indeed, increasingly accurate predictions, computing power coupled with the astonishing progress of DL could uncritically reopen the way to a sort of new positivism based on data, both scientific and social, driven by factors external to science rather than by scientists - who are well aware of what is at stake in the scientific method. 
We can certainly talk about a real paradigm shift in Kuhn's sense, not towards a new positivism based on the objectivity of the method, but rather on the presumed neutrality of algorithms and on the possibility of explaining everything through data, without any predictive model (in science) and without the need of imagine a better one for the society. 
If we continue to endorse Anderson's view, we will go in the direction of a data-dominated development, a dystopia where the status quo is reproduced in vicious circles. 
So it is also up to the scientific community to restate the problem by re-contextualising science itself and helping to deconstruct its aura of absolute objectivity, neutrality and a-historicity that today characterises any debate around data and algorithms. This work is a step in this direction.

\vspace{0.5cm}
\noindent
{\bf Acknowledgements:}
It is a pleasure to thank Sharu Trevisiol for the translation.


\bibliographystyle{utphys}
\bibliography{biblio_ML_2}

\end{document}